\documentclass[lettersize,journal]{IEEEtran}
\usepackage{cite}
\usepackage{amsmath,amsfonts}
\usepackage{algorithmic}
\usepackage{array}
\usepackage[caption=false,font=normalsize,labelfont=sf,textfont=sf]{subfig}
\usepackage{textcomp}
\usepackage{stfloats}
\usepackage{url}
\usepackage{verbatim}
\usepackage{graphicx}
\usepackage{balance}
\usepackage{graphicx}
\usepackage{xcolor}
\usepackage{tikz}
\usepackage[caption=false]{subfig}
\usetikzlibrary{shapes.geometric,arrows.meta,positioning}

\def\BibTeX{{\rm B\kern-.05em{\sc i\kern-.025em b}\kern-.08em
    T\kern-.1667em\lower.7ex\hbox{E}\kern-.125emX}}

\begin{document}
\title{Structural Decoupling and Current-Angle Steering for Post-Fault Recovery of Current-Limited Grid-Forming Inverters}
\author{
	Neethu Sajeev,~\IEEEmembership{Graduate Student Member,~IEEE,}
	Stephen Arinze Obi,~\IEEEmembership{Graduate Student Member,~IEEE,}
	and Jae-Jung Jung,~\IEEEmembership{Senior Member,~IEEE}
}
\markboth{}%
{Structural Decoupling and Candidate-Based Current-Angle Steering for Post-Fault Recovery of Grid-Forming Inverters Under Current Limitation}

\maketitle

\begin{abstract}
	Reliable fault recovery of grid-forming (GFM) converters under current-limited conditions is increasingly important as inverter based resources replace synchronous generation. Existing current-limiting strategies primarily focus on current-angle regulation and synchronization trajectory shaping, while the interaction between the current limiter and the voltage control structure remains insufficiently understood. Consequently, post-fault recovery may exhibit converter trapping in current-limited control (CLC) or oscillatory transitions between CLC and constant voltage control (CVC).
	
	This paper shows that, under conventional PI-based voltage control, the interaction between the voltage controller and the current limiter creates a moving recovery boundary that contributes to these recovery failures. To address this issue, a post-fault recovery framework is proposed that combines structurally decoupled virtual admittance voltage control with current-angle steering. The proposed framework simultaneously improves synchronization trajectory evolution and stabilizes the recovery boundary during fault recovery. Experimental validation on a 3-kVA GFM inverter prototype confirms reliable post-fault synchronization recovery under both symmetrical and unsymmetrical voltage sag conditions, with trapping and oscillatory CLC-CVC transitions eliminated.
\end{abstract}
\section{Introduction}

\IEEEPARstart{T}{he} increasing penetration of inverter based resources (IBRs) is fundamentally reshaping modern power systems \cite{Hatziargyriou,Christensen}. As synchronous generators are progressively displaced, maintaining synchronization stability under reduced inertia and weak grid conditions has become increasingly challenging \cite{Zhang}. In addition, the limited fault current capability of power electronic converters introduces new challenges in protection and transient performance during grid disturbances \cite{Fan}. In this context, grid-forming (GFM) converters have attracted significant attention due to their ability to establish voltage and frequency while actively participating in system dynamics \cite{Wang,Harnefors}. Unlike grid-following converters, GFM units operate as controlled voltage sources whose fault dynamics directly influence synchronization stability \cite{Gao,Salem}. Consequently, reliable operation of GFM converters under severe disturbances has become a critical requirement for future power systems.

A major challenge in GFM converters is their limited over current capability during severe grid disturbances \cite{Narula}. Since power electronic converters cannot sustain large fault currents for extended durations, current limiting control is required to ensure safe operation and fault ride through capability \cite{Bottrell,ZhangGFM}. Existing current limiting approaches are broadly categorized into virtual impedance methods and current reference saturation techniques \cite{Paquette,FanReview}. Virtual impedance methods regulate current indirectly through voltage modification but may suffer from transient overcurrent due to limited control bandwidth \cite{Qoria,Xiong}. In contrast, current reference saturation methods provide fast and direct current limiting but fundamentally modify the converter voltage source characteristics and synchronization dynamics \cite{Kay}. Moreover, converter network interactions significantly influence stability during current-limited operation \cite{RiosCastro,Wang2025}. Therefore, current limitation should be regarded not only as a protection mechanism, but also as a dominant factor governing the large signal dynamics of GFM converters.

With the inclusion of current limiting control, post-fault recovery of GFM converters has emerged as a distinct stability problem. Previous studies have shown that converters may fail to return to normal voltage regulation after fault clearance, resulting in trapping in current-limited control (CLC) or oscillatory transitions between CLC and constant voltage control (CVC) \cite{Zhan,FanFault,Yang2025,YangGraphical}. In addition, current saturation modifies synchronization characteristics and may introduce complex transient phenomena such as backward swing dynamics \cite{Rokrok,Xue}. These observations indicate that current limitation fundamentally alters post-fault synchronization behavior.

To analyze these phenomena, trajectory based and equilibrium based approaches have been widely adopted. Existing studies have shown that the injected current-angle strongly influences synchronization trajectories and transient stability during current-limited operation \cite{ArjomandiNezhad,FanFault,Rokrok}. Consequently, fixed angle and priority based current injection strategies have been proposed to improve fault recovery performance. However, many of these methods rely on offline selection of the injected current-angle using grid-dependent information such as grid impedance, operating condition, or fault characteristics, which may limit robustness under varying grid scenarios and parameter uncertainties \cite{FanFault,Rokrok,Yang2025}. Moreover, most existing approaches primarily focus on synchronization trajectory shaping while implicitly assuming ideal voltage control dynamics. Current saturation strategies have also been extensively studied for low voltage ride through enhancement and reactive power support, whereas their interaction with post-fault recovery dynamics remains insufficiently addressed \cite{Pal,Fernandez}.

Despite these developments, a fundamental limitation remains unresolved. Existing studies largely neglect the structural coupling between the current limiter and the voltage control loop. In practical cascaded control implementations, the voltage controller generates the current reference, which is subsequently constrained by the limiter, producing coupled dynamics that deviate from ideal assumptions \cite{Baeckeland}. 
Previous studies have shown that current saturation may introduce windup related recovery difficulties and motivate anti-windup compensation \cite{Hu,Zhuang}. However, recent studies have reported that trapping in CLC and oscillatory CLC-CVC transitions may still occur even when anti-windup measures are employed \cite{YangGraphical,Yang2025}. These observations suggest that post-fault recovery is influenced not only by windup phenomena but also by the interaction between synchronization trajectory evolution, current-limited operation, and recovery boundary dynamics.

To address this issue, this paper investigates the coupled dynamics between current limiting mechanisms and voltage control structures in GFM converters. The analysis reveals that conventional PI-based voltage regulation introduces a moving switching boundary mechanism during current-limited operation due to the interaction between the voltage controller and the saturated converter current. Based on this observation, a post-fault recovery framework combining structurally decoupled virtual admittance voltage regulation and current-angle steering is proposed. The proposed framework simultaneously regulates the synchronization trajectory and stabilizes the recovery boundary dynamics, thereby suppressing trapping and oscillatory recovery behavior. Experimental validation under both symmetrical and unsymmetrical voltage sag conditions confirms reliable post-fault synchronization recovery under varying grid conditions.

Existing current-angle based recovery methods mainly utilize the injected current-angle to reshape synchronization trajectories and enlarge admissible recovery regions during current-limited operation \cite{FanFault,Yang2025}, while implicitly assuming ideal voltage control dynamics. References \cite{Hu,Zhuang} identify integrator windup as a source of recovery difficulty and propose anti-windup compensation. However, as demonstrated experimentally in this paper, trapping and oscillatory CLC--CVC transitions persist even when anti-windup is applied, indicating that the recovery problem is not fully resolved by windup suppression alone. The present work identifies that the root cause is a moving switching boundary introduced by the structural coupling between the PI voltage controller and the current limiter, which remains active even under anti-windup operation. Unlike these prior approaches, the proposed framework eliminates the coupling mechanism at its source through a structurally decoupled virtual admittance voltage controller, while candidate-based current-angle steering shapes the recovery trajectory without requiring grid-angle measurement. This combination constitutes the primary novelty of the proposed framework.

The main contributions of this work are summarized as follows:
\begin{itemize}
	\item Recovery oriented characterization of the coupled interaction between synchronization trajectory evolution and CLC/CVC transition dynamics under current-limited operation.
	
	\item Development of a reduced-order framework linking PCC voltage trajectory evolution, synchronization dynamics, and CLC/CVC switching behavior during current-limited operation.
	
	\item Proposal of a post-fault recovery framework combining structurally decoupled virtual admittance voltage regulation with current-angle steering.
	
	\item Experimental validation under symmetrical and unsymmetrical voltage sag conditions demonstrating suppression of trapping and oscillatory recovery behavior.
\end{itemize}

The remainder of this paper is organized as follows. Section II presents the system model and control structure. Section III analyzes the interaction between current limitation and voltage control. Section IV introduces the proposed fault recovery enhancement framework. Section V presents experimental validation results under symmetrical and unsymmetrical voltage sag conditions. Finally, Section VI concludes the paper.
\begin{figure}[!t]
	\centering
	\includegraphics[width=3.3in]{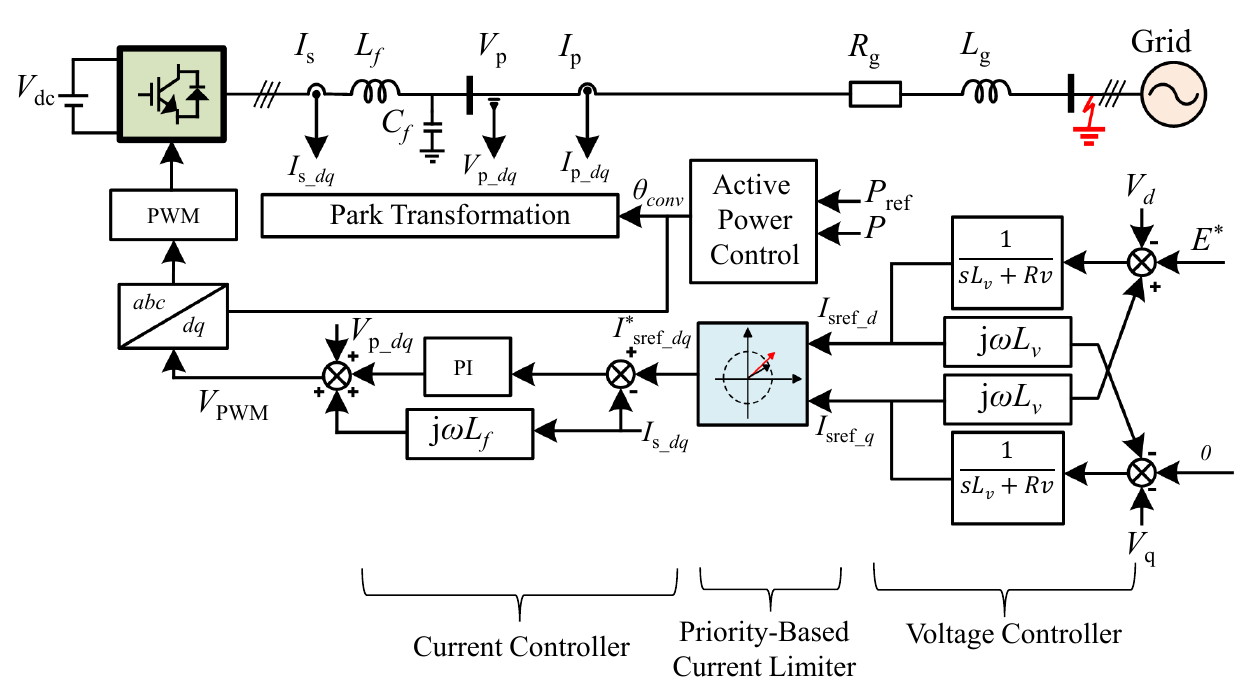}
	\caption{Control structure of the grid-connected GFM inverter with the proposed priority based current limiting scheme.}
	\label{fig1}
\end{figure}
\section{System Configuration and Control Structure}

\subsection{System Description}

The overall control structure of the studied grid-connected grid-forming (GFM) inverter equipped with the proposed current limiting framework is illustrated in Fig.~\ref{fig1}. The inverter is connected to the utility grid through an LC filter and a grid-side impedance represented by \(R_g\) and \(L_g\). The converter operates as a controlled voltage source and regulates the PCC voltage through an outer active power control (APC) loop and an inner voltage control loop.

The APC loop generates the converter angular frequency \(\omega_{conv}\), which is subsequently integrated to obtain the converter synchronization-angle \(\theta_{conv}\). The PCC voltage and converter current are transformed into the synchronous \(dq\) reference frame through the Park transformation. Based on the PCC voltage regulation error, the voltage control loop generates the converter current reference \(I^{ref}_{sdq}\).

During severe grid disturbances, the demanded converter current may exceed the allowable converter current capability. Under such conditions, the current limiter becomes active and constrains the converter current magnitude according to the predefined current limit. Consequently, the converter transitions from CVC to CLC, where synchronization dynamics become strongly influenced by the injected current-angle.

\subsection{Control Structure and Current Limiting Strategy}

For fault recovery analysis, the studied GFM inverter employs an active power control (APC) loop together with an inner voltage control loop. The converter frequency is generated according to the active power mismatch as
\begin{equation}
	\omega_{conv}=\omega_{ref}+K_{PLC}G_{LPF}(s)\left(P_{ref}-P_{act}\right)
	\label{eq1}
\end{equation}
where \(\omega_{ref}\) is the nominal grid angular frequency, \(P_{ref}\) and \(P_{act}\) are the reference and measured active power, respectively, and \(K_{PLC}\) denotes the APC gain. The low-pass filter transfer function is represented by \(G_{LPF}(s)\).
During severe voltage sag conditions, the converter apparent power capability is reduced due to current limitation and reactive power support requirements. The available apparent power is scaled according to the PCC voltage magnitude \(V_{pu}\) as
\begin{equation}
	S_{new}=V_{pu}S_n
	\label{eq2}
\end{equation}
where \(S_n\) is the rated converter apparent power. Based on the sag depth, the active and reactive power references are modified as

\begin{equation}
	\begin{split}
		P_{ref,new} &=
		\begin{cases}
			P_{ref}, & V_{pu}>0.9,\\
			\sqrt{S_{new}^{2}-Q_{ref,new}^{2}}, & 0.5<V_{pu}\leq0.9,\\
			0, & V_{pu}\leq0.5,
		\end{cases}
		\\
		Q_{ref,new} &=
		\begin{cases}
			Q_{ref}, & V_{pu}>0.9,\\
			2S_{new}(1-V_{pu}), & 0.5<V_{pu}\leq0.9,\\
			S_{new}, & V_{pu}\leq0.5.
		\end{cases}
	\end{split}
	\label{eq3}
\end{equation}
To prevent excessive synchronization acceleration during severe faults, the APC gain is adaptively modified as
\begin{equation}
	K_{PLC,ada}=
	\frac{K_{PLC}}
	{1+\alpha\left(\dfrac{(P_{ref}-P_{act})^2}{P_{ref,new}^2+\epsilon_P}\right)}
	\label{eq4}
\end{equation}
where $\alpha$ is a tuning coefficient and $\epsilon_P$ is a small positive constant introduced to avoid singularity when $P_{ref,new}=0$ under deep voltage sag conditions.

During normal operation, the converter current reference generated by the voltage controller is directly applied to the modulation stage. However, during severe disturbances, the demanded converter current may exceed the allowable current limit. Therefore, a priority based current limiter is employed to constrain the converter current magnitude within the predefined current capability \(I_{s,\max}\). The limited current reference is expressed as
\begin{equation}
	\bar{I}^{*}_{sref,dq}=
	\begin{cases}
		I_{s,\max}e^{j\phi}, & \left\|I_{sref,dq}\right\|>I_{s,\max} \\[4pt]
		I_{sref,dq}, & \left\|I_{sref,dq}\right\|\leq I_{s,\max}
	\end{cases}
	\label{eq5}
\end{equation}
where \(\bar{I}^{*}_{sref,dq}\) denotes the saturated current reference and \(\phi\) represents the injected current-angle during current-limited operation. To prevent integrator windup during current saturation, the voltage controller integrator gain is set to zero when the current limiter is active, effectively freezing the integrator state during current-limited operation.

Since the inner voltage and current control dynamics evolve at a significantly faster timescale than the APC dynamics, the fast electrical transients can be neglected for fault recovery analysis \cite{Wang,YangGraphical}. Under current-limited operation, the converter current magnitude remains constrained by the limiter while the injected current-angle remains controllable. Consequently, the synchronization and post-fault recovery characteristics are primarily governed by the interaction between the APC dynamics, the current limiting mechanism, and the voltage control structure. Based on these considerations, a reduced-order framework is developed in the following section to analyze the fault recovery dynamics of the studied GFM inverter under current limitation.
\section{Dynamic Interaction Between Current Limitation and Voltage Control}

This section investigates the fault recovery dynamics of the studied GFM inverter under current-limited operation. First, a reduced-order representation is developed to describe the dominant synchronization behavior during fault conditions. Based on this framework, the PCC voltage trajectory and the switching characteristics between CLC and CVC are analyzed. Subsequently, the interaction between the current limiter and the voltage control structure is examined to explain the occurrence of trapping and oscillatory recovery phenomena during post-fault operation.

\subsection{Reduced-Order Representation of Current-Limited Operation}

To capture the dominant fault recovery dynamics, the fast switching, filter, and inner current control dynamics are neglected. This assumption is commonly adopted in large signal stability studies of GFM converters since the outer synchronization dynamics evolve at a significantly slower timescale than the inner control loops \cite{FanFault,Rokrok}.

During current-limited operation, the converter current magnitude is constrained by the current limiter while the injected current-angle remains controllable. Consequently, the converter current phasor can be expressed as
\begin{equation}
	I_s = I_{s,\max} e^{j\phi}
	\tag{6}
\end{equation}
where \(I_{s,\max}\) represents the converter current limit, and \(\phi\) specifies the current-injection angle applied during saturation \cite{FanFault,Rokrok}.
Let \(\delta\) denote the converter grid angle difference. In the converter reference frame, the grid voltage is written as
\begin{equation}
	V_g = V_g e^{-j\delta}
	\tag{7}
\end{equation}
Assuming a predominantly inductive grid impedance, the PCC voltage during current-limited operation is expressed as

\begin{equation}
	V_{pcc}=V_g e^{-j\delta}+jX_g I_{s,\max} e^{j\phi}
	\tag{8}
\end{equation}

where \(X_g=\omega_gL_g\) is the equivalent grid reactance. Equation (8) indicates that the PCC voltage is jointly determined by the synchronization-angle \(\delta\) and the injected current-angle \(\phi\). The injected current-angle $\phi$ is defined following the convention adopted in \cite{YangGraphical,Yang2025}, where $\phi \in [-\pi/2,0]$. Consequently, the phasor diagrams in Figs.~2--3 depict the geometric angle $-\phi$, while the analytical expressions are written in terms of the control variable $\phi$. The grid impedance is assumed to be predominantly inductive in the reduced-order analysis, i.e., \(X_g \gg R_g\). Under this approximation, the resistive voltage drop is neglected to obtain the compact geometric representation in (8).

Expressing \(V_{pcc}=V_{pd}+jV_{pq}\), (8) can be rearranged as
\begin{equation}
	\left(V_{pd}+X_g I_{s,\max}\sin\phi\right)^2+
	\left(V_{pq}-X_g I_{s,\max}\cos\phi\right)^2
	=V_g^2
	\tag{9}
\end{equation}

which represents a circular PCC voltage trajectory in the phasor plane during current-limited operation \cite{Yang2025,Xue}. This geometric relationship forms the basis for the subsequent synchronization and recovery analysis.
The active power transferred during current-limited operation is expressed as
\begin{equation}
	P=\operatorname{Re}\{V_{pcc}I_s^{*}\}
	\tag{10}
\end{equation}
Substituting (6) and (8) into (10) yields
\begin{equation}
	P(\delta,\phi)=V_gI_{s,\max}\cos(\delta+\phi)
	\tag{11}
\end{equation}
which indicates that the saturated power transfer depends on the relative angle between the synchronization-angle and the injected current vector \cite{Rokrok}.
Since the APC dynamics are governed by the active power mismatch, the reduced-order angle dynamics under current limitation are expressed as
\begin{equation}
	\dot{\delta}
	=
	K_{PLC,ada}G_{LPF}(s)
	\left(P_{ref,new}-P(\delta,\phi)\right)
	\tag{12}
\end{equation}
Equations (8)–(12) establish the reduced-order framework used in the following subsections to investigate the PCC voltage trajectory, switching characteristics, and post-fault synchronization behavior of the studied GFM inverter under current limitation.

\subsection{PCC Voltage Trajectory Under Current Limitation}

During current-limited operation, the PCC voltage trajectory becomes dependent on both the synchronization-angle and the injected current vector. Unlike normal CVC operation, where the PCC voltage is directly regulated by the voltage controller, the PCC voltage under current saturation evolves according to the constrained current injection and the converter-grid angle difference \cite{FanReview,Yang2025}. From (9), the center of the PCC voltage trajectory is given by
\begin{equation}
	\left(
	-X_gI_{s,\max}\sin\phi,\;
	X_gI_{s,\max}\cos\phi
	\right)
	\tag{13}
\end{equation}
while the trajectory radius remains equal to the grid voltage magnitude \(V_g\). Consequently, variation of the injected current-angle shifts the PCC voltage trajectory in the phasor plane and modifies the associated synchronization behavior during fault conditions.

According to (11), the synchronization dynamics are governed by the relative angle between the converter synchronization-angle and the injected current vector. Linearizing (11) around the equilibrium point $\delta^\ast$ yields

\begin{equation}
	\Delta P = -V_gI_{s,\max}\sin(\delta^\ast+\phi)\Delta\delta.
	\tag{14}
\end{equation}
 Equation (14) shows that the sign of $-\sin(\delta^\ast+\phi)$ determines the local synchronization characteristic. Positive values correspond to restoring behavior, where the resulting electrical power variation opposes the synchronization-angle deviation. Conversely, negative values correspond to anti-restoring behavior, where the electrical power variation reinforces the synchronization-angle deviation. Neglecting the LPF dynamics for qualitative interpretation, the synchronization evolution can be approximated as
\begin{equation}
	\dot{\delta}
	\approx
	K_{PLC,ada}
	\left(
	P_{ref,new}
	-
	V_g I_{s,\max}\cos(\delta+\phi)
	\right).
	\tag{15}
\end{equation}
Under the restoring condition, the resulting electrical power variation acts to reduce the synchronization-angle deviation,
thereby promoting recovery toward the equilibrium point.
Conversely, operation within the anti-restoring region reinforces
the synchronization-angle deviation and may drive the trajectory
away from the equilibrium point. Therefore, the injected current-angle directly influences the recovery characteristics of the
saturated power angle dynamics during fault recovery.

These observations indicate that the injected current-angle directly determines the effective restoring region during current-limited operation. Under identical fault conditions, different current injection angles may produce substantially different synchronization trajectories and post-fault recovery responses. Certain trajectories evolve toward recoverable operating regions, whereas others remain trapped in current-limited operation or repeatedly approach the switching boundary between CVC and CLC operation.
\begin{figure}[!t]
	\centering
	\includegraphics[width=3.1in]{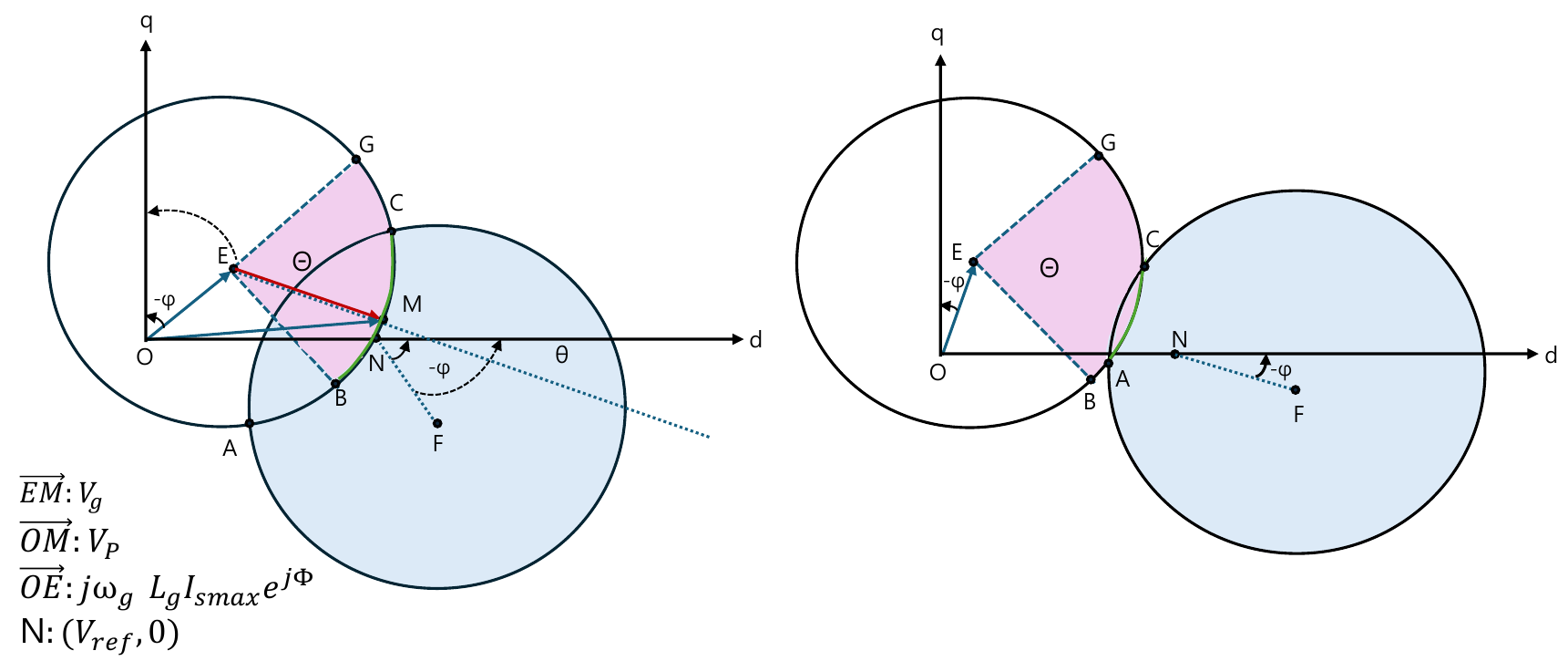}
	\caption{Geometric interpretation of the switching 
		boundary under PI-based voltage control. Arc 
		$\widehat{BC}$ represents the successful recovery 
		region; arc $\widehat{AB}$ corresponds to the 
		oscillatory transition region; the remaining arc 
		$\widehat{AC}$ (through the left semicircle) denotes 
		the trapping region; $G$ marks the boundary of 
		sector $\Theta$; and $F$ is the center of the 
		switching boundary disk.}
	\label{fig:fig2}
\end{figure}

\begin{figure}[!t]
	\centering
	\includegraphics[width=3.1in]{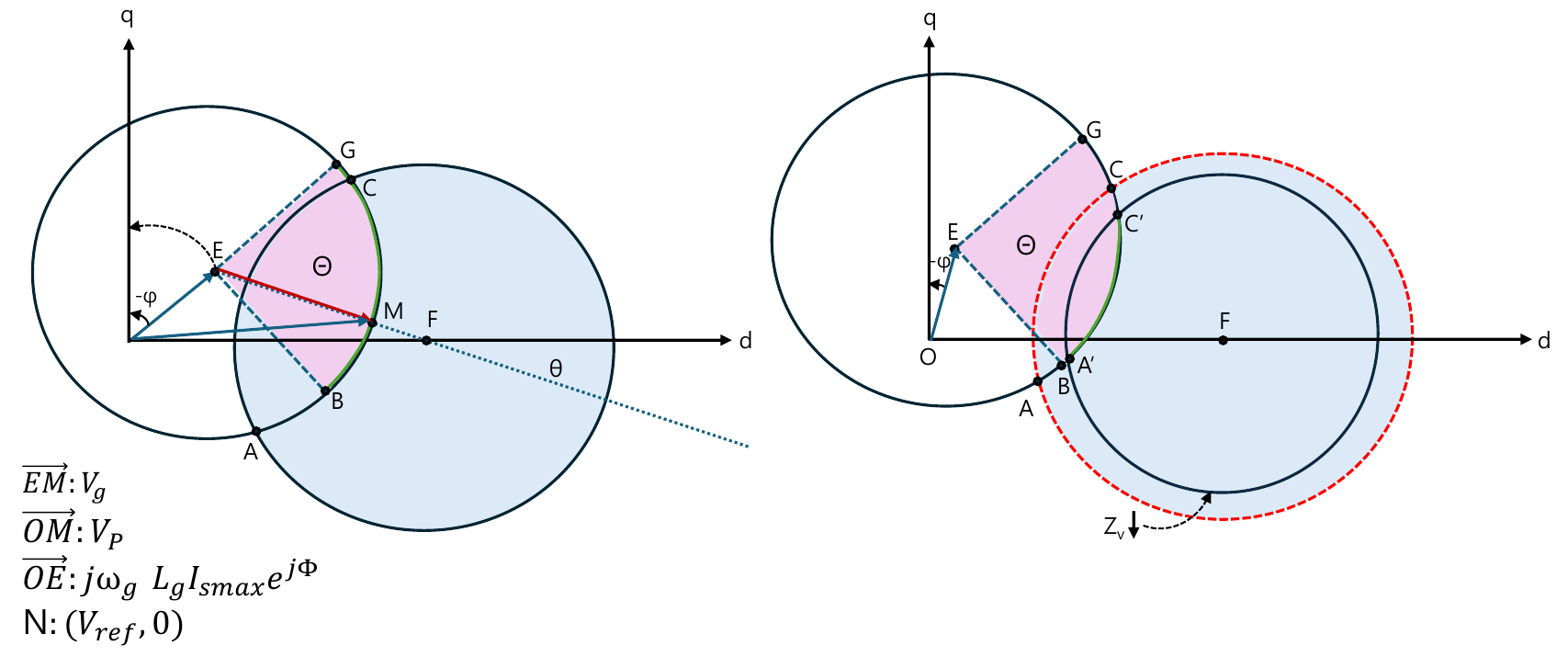}
	\caption{Geometric interpretation of the switching boundary under virtual-admittance voltage control. Compared with Fig.~2, the structural decoupling contracts the switching boundary disk $F$ toward $N=(V_{ref},0)$, shifting the recovery arc from $\widehat{BC}$ to the enlarged region $\widehat{A^{\prime}C^{\prime}}$, such that the entire arc $\widehat{AC}$ lies within sector $\Theta$, eliminating oscillatory transitions and enabling consistent CLC-CVC recovery.}
	\label{fig:fig3}
\end{figure}
During normal operation, the converter operates in CVC mode, where the PCC voltage is regulated according to the voltage reference. Under severe disturbances, however, the demanded converter current may exceed the allowable current capability, activating the current limiter and forcing the converter into CLC operation \cite{FanFault,Yang2025}.

Let \(I^{ref}_{sdq}\) denote the unconstrained current reference generated by the voltage controller. The converter enters CLC operation when
\begin{equation}
	\left\| I^{ref}_{sdq} \right\| > I_{s,\max}
	\tag{16}
\end{equation}
and returns to CVC operation when
\begin{equation}
	\left\| I^{ref}_{sdq} \right\| \leq I_{s,\max}.
	\tag{17}
\end{equation}

Accordingly, the ideal switching boundary between CVC and CLC operation is defined as
\begin{equation}
	\left\| I^{ref}_{sdq} \right\| = I_{s,\max}.
	\tag{18}
\end{equation}

Under ideal conditions, the transition depends only on the intersection between the voltage controller current demand and the allowable current boundary. Therefore, once the demanded current decreases below the current limit after fault clearance, the converter is expected to naturally return to CVC operation.

In practical cascaded control implementations, however, the switching behavior is additionally influenced by the interaction between the voltage control loop and the current limiter. During current saturation, the voltage controller continues responding to the PCC voltage error while the converter current magnitude remains constrained by the limiter. Consequently, the internally generated current reference may deviate significantly from the actual saturated converter current.
The unconstrained current generated by the voltage controller is defined as
\begin{equation}
	I_{soft}=I^{ref}_{sdq}
	\tag{19}
\end{equation}
while the saturated converter current is represented as
\begin{equation}
	I_{sat}=I_{s,\max}e^{j\phi}.
	\tag{20}
\end{equation}

During severe disturbances, the mismatch between \(I_{soft}\) and \(I_{sat}\) may become substantial due to accumulated voltage controller action.
 As a result, the effective transition between CVC and CLC operation depends not only on the instantaneous current magnitude in (18), but also on the internal voltage controller dynamics.

Under conventional PI-based voltage regulation, the effective switching boundary becomes strongly dependent on the injected current-angle and transient controller dynamics, as illustrated in Fig.~\ref{fig:fig2}. The shaded disk centered at $F$ represents the effective CLC/CVC recovery boundary generated by the interaction between the voltage controller and the current limiter during current-limited operation. Although the integrator is frozen during current saturation, the proportional voltage-error response and the retained integral state still cause $I_{soft}$ to deviate from $I_{sat}$, thereby shifting the effective CLC-CVC switching boundary and distorting the recovery characteristics even with anti-windup.

In contrast, under virtual admittance based voltage regulation, the demanded current remains directly proportional to the PCC voltage error without integral accumulation. Consequently, the mismatch between the unconstrained and saturated current is significantly reduced, resulting in comparatively stable recovery boundary characteristics, as illustrated in Fig.~\ref{fig:fig3}. The shaded angular region corresponds to the admissible synchronization-angle sector \(\Theta\), while the shaded disk represents the effective switching boundary region during current-limited operation. The virtual admittance parameter further influences the recovery boundary geometry. Increasing the virtual admittance enlarges the corresponding boundary region and modifies the transition behavior during fault recovery. Therefore, excessively small virtual admittance values may provide insufficient recovery boundary stabilization, whereas excessively large values may degrade voltage regulation performance. In practice, moderate virtual admittance selection provides an effective tradeoff between recovery boundary stabilization and voltage regulation dynamics.

The admissible synchronization-angle sector for avoiding immediate re-entry into CLC after recovery is obtained from the CVC current limit condition
\begin{equation}
	\left|\frac{V_{ref}-V_g e^{-j\delta}}{jX_g}\right|
	\leq I_{s,\max}
	\tag{21}
\end{equation}
Equivalently,
\begin{equation}
	\cos\delta \geq \eta,
	\qquad
	\eta=
	\frac{
		V_{ref}^{2}+V_g^{2}-X_g^{2}I_{s,\max}^{2}
	}{
		2V_{ref}V_g
	}
	\tag{22}
\end{equation}

Thus, the admissible non-reentry region is given by
\begin{equation}
	\delta \in \Theta =
	[-\cos^{-1}(\eta),\,\cos^{-1}(\eta)]
	\tag{23}
\end{equation}
within one \(2\pi\) period.

Although the admissible sector \(\Theta\) is determined by the CVC current limit condition and therefore remains applicable to both conventional and proposed structures, the proposed virtual admittance based voltage control framework significantly reduces the sensitivity of the exit boundary dynamics to \(\phi\). Consequently, the recovery trajectory interacts with \(\Theta\) more consistently during current-angle steering.

Under severe disturbance conditions, the converter may remain trapped in current-limited operation even after the synchronization trajectory re-enters a theoretically recoverable region. In addition, repeated interaction with the switching boundary may produce oscillatory transitions between CLC and CVC operation before the converter reaches a stable post-fault operating condition.
\subsection{Dynamic Implication of Exit Boundary Coupling}

The preceding analysis shows that post-fault recovery of current-limited GFM converters is governed by both synchronization-trajectory evolution and recovery boundary dynamics. Under conventional PI-based voltage regulation, the effective recovery boundary varies during current-limited operation because the internally generated current reference continues evolving independently of the saturated converter current. Consequently, the exit condition from CLC depends not only on the synchronization trajectory but also on the internal voltage-controller state, which may result in delayed recovery or oscillatory CLC-CVC transitions.

In contrast, the proposed virtual admittance based voltage control structure maintains structural decoupling between the unconstrained soft current and the saturated converter current, thereby reducing the sensitivity of the recovery boundary to injected current-angle variation and transient voltage-controller dynamics. As a result, the injected current-angle primarily governs the synchronization trajectory, while the recovery boundary remains comparatively consistent with respect to the admissible synchronization-angle sector $\Theta$. Therefore, reliable post-fault recovery requires both favorable synchronization trajectory shaping and stabilization of the recovery boundary dynamics. Based on these observations, the proposed fault recovery enhancement framework is introduced in the following section.

\section{Proposed Fault Recovery Enhancement Framework}

Based on the preceding analysis, a post-fault recovery framework is developed by combining structurally decoupled virtual admittance voltage control with current-angle steering. The proposed framework simultaneously regulates synchronization trajectory evolution and improves recovery boundary consistency during current-limited operation. This section introduces the candidate-based current-angle steering mechanism, followed by the online recovery logic and the resulting dynamic recovery characteristics.

\subsection{Candidate-Based Recovery Trajectory Regulation}

During current-limited operation, the synchronization behavior of the GFM inverter is strongly influenced by the injected current-angle. As indicated by the saturated power-angle relationship in (11), changing the injected current-angle modifies the synchronization trajectory and the local restoring characteristic during fault recovery. Based on this observation, the proposed strategy evaluates a finite set of candidate current-angles online and selects the angle that provides favorable recovery behavior while avoiding excessive synchronization acceleration.

For practical implementation, direct use of the synchronization-angle $\delta$ is avoided because it requires explicit grid-angle information. Instead, a PLL-free PCC voltage-angle proxy is obtained from the measured PCC voltage components in the converter reference frame as
\begin{equation}
	\delta_p=\mathrm{atan2}(v_q,v_d)
	\tag{24}
\end{equation}
where $v_d$ and $v_q$ denote the PCC voltage components transformed into the converter reference frame using the converter synchronization-angle $\theta_{conv}$ generated internally by the APC loop, as shown in Fig.~\ref{fig1}. Since $\theta_{conv}$ is available without a PLL, $\delta_p$ is computed directly from measured PCC voltage without grid-angle estimation. The proxy angle $\delta_p$ captures the instantaneous orientation of the PCC voltage trajectory during current-limited operation and is used only for online candidate evaluation.

The candidate evaluation function is defined as
\begin{equation}
	J(\phi_i)
	=
	\lambda \max\left(0,\sin(\delta_p-\phi_i)\right)^2
	+
	\rho(\phi_i-\phi_{\mathrm{prev}})^2 
	\tag{25}
\end{equation}

The first term penalizes non-restoring candidate angles, while the second term suppresses excessive variation of the injected current-angle during steering transitions. Defining $s_{r,i}=-\sin(\delta_p-\phi_i)$, a positive value of $s_{r,i}$ indicates restoring behavior, whereas a negative value indicates non-restoring behavior. Therefore, the penalty term $\max(0,\sin(\delta_p-\phi_i))^2=\max(0,-s_{r,i})^2$ penalizes only candidates with negative recovery margin. The quantity $s_{r,i}$ is distinct from the analytical synchronization slope $-\sin(\delta+\phi)$ in (14), and serves as a locally measurable PCC-frame recovery indicator. Since both $\delta_p$ and $\phi_i$ are expressed in the converter reference frame, $s_{r,i}$ is used to rank candidate angles without requiring a PLL or explicit grid-angle estimation. Here, $\lambda$ and $\rho$ are weighting coefficients, and $\phi_{\mathrm{prev}}$ denotes the previously applied current-angle.

 The candidate evaluation is performed over $N_{\mathrm{cand}}=11$ candidate current-angles uniformly distributed around the nominal current-angle and constrained to the admissible steering interval. To improve robustness against measurement noise and transient voltage disturbances, the computed proxy angle is low-pass filtered during steering operation before being supplied to the candidate evaluation process. Candidate angles violating feasibility constraints, including the SEP margin $\Delta_{\mathrm{SEP}}$, are discarded during the evaluation process. 
 The weighting coefficients are chosen as $\lambda=8$ and $\rho=0.2$, while the hysteresis thresholds are set to $s_{\mathrm{low}}=0.02$ and $s_{\mathrm{crit}}=0.05$. To avoid unnecessary steering action under deep voltage-sag conditions, steering is temporarily disabled when the PCC voltage magnitude falls below $V_{\mathrm{deep}}=0.60$~p.u. Consequently, the selected current-angle improves the recovery trajectory while limiting excessive synchronization acceleration near the switching boundary. Accordingly, the candidate selection process utilizes a locally measurable implementation proxy rather than direct evaluation of the analytical synchronization slope.

\begin{figure}[!t]
	\centering
	\includegraphics[width=3.1in]{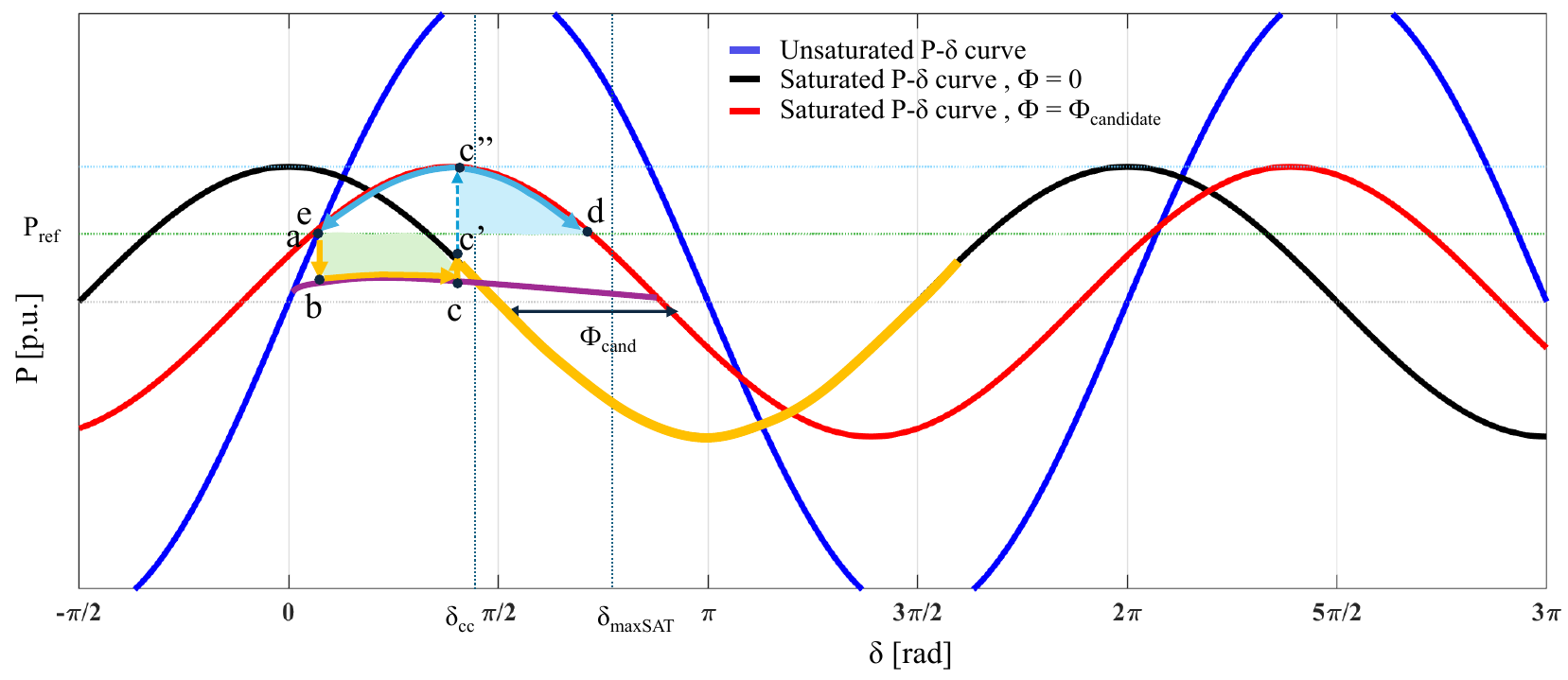}
	\caption{Operation mechanism of GFM under voltage sag with candidate-based current-angle steering.}
	\label{fig:pd_candidate}
\end{figure}

Fig.~\ref{fig:pd_candidate} illustrates the operating principle of the proposed candidate-based steering strategy. Points $c'$ and $c''$ denote the post-fault operating points on the saturated $P$--$\delta$ curves corresponding to $\phi=0$ and $\phi=\phi_{\mathrm{cand}}$, respectively, illustrating how the candidate angle modifies the saturated power-angle characteristic and the resulting recovery trajectory. By adaptively adjusting the injected current-angle according to the instantaneous recovery condition, the synchronization trajectory is steered toward favorable restoring regions. Furthermore, because the unconstrained soft current remains structurally decoupled from the saturated converter current, the trajectory shaping effect of the injected current-angle can be utilized without significantly altering the recovery boundary dynamics. Consequently, the proposed strategy improves synchronization recovery while reducing oscillatory CLC-CVC transitions.

\subsection{Online Recovery Control Logic}

\begin{figure}[!t]
	\centering
	\includegraphics[width=2.4in]{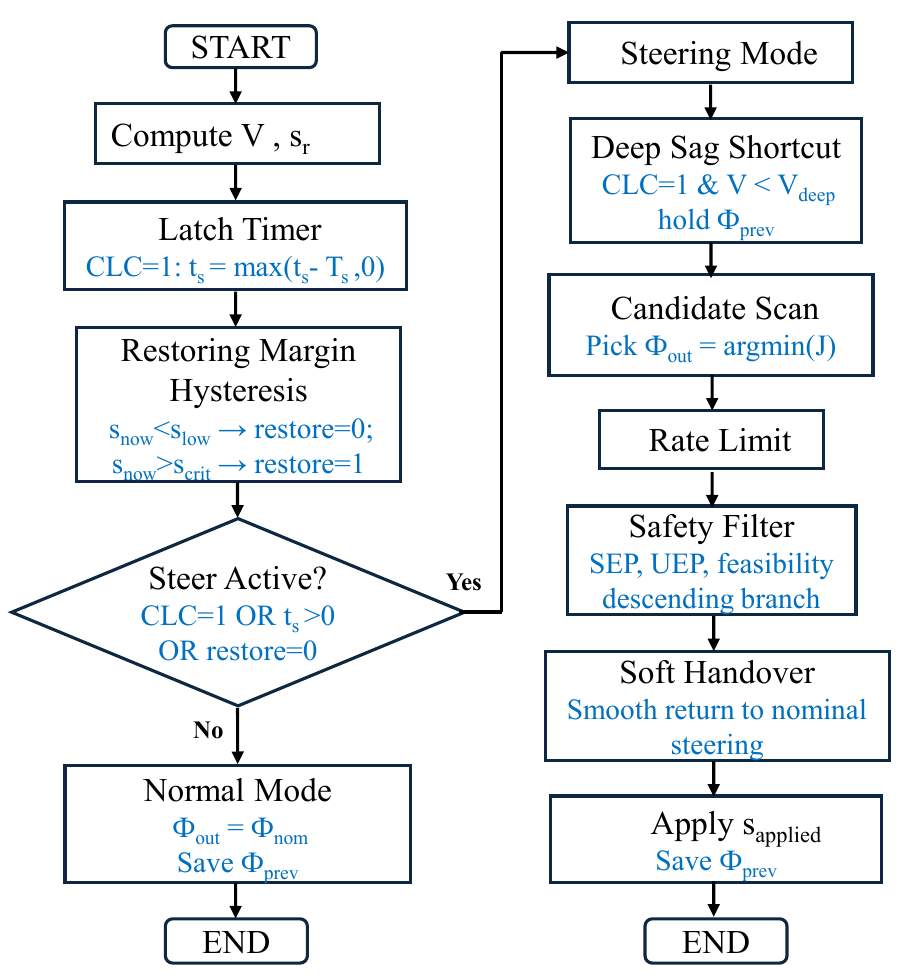}
	\caption{Operational flowchart of the proposed candidate-based current-angle steering and recovery framework.}
	\label{fig:fig5}
\end{figure}

To achieve stable fault recovery under varying disturbance conditions, the proposed framework employs an online recovery mechanism that coordinates fault detection, candidate-angle evaluation, and transition management during current-limited operation, as illustrated in Fig.~\ref{fig:fig5}. During normal operation, the converter operates in CVC mode with the nominal current-angle. Once a fault is detected, the controller enters CLC operation and activates candidate-based current-angle steering. Here, $s_r=-\sin(\delta_p-\phi)$ denotes the recovery indicator computed from the PCC voltage angle $\delta_p=\mathrm{atan2}(v_q,v_d)$. The quantities $s_{now}$ and $s_{applied}$ denote the indicator evaluated using the previously applied angle $\phi_{prev}$ and the selected output angle $\phi_{out}$, respectively, and $t_s$ is the steering hold timer.

During fault and post-fault recovery, candidate current-angles are evaluated within $-\pi/2 \le \phi \le 0$; the selector may choose $\phi=0$ during the sag when this candidate minimizes (25), which is consistent with maximum saturated active-power transfer. After fault clearance, $\phi$ is updated according to the measured recovery condition to obtain a favorable restoring tendency with limited synchronization acceleration. The adaptive APC gain introduced in Section II further moderates synchronization-angle acceleration during severe voltage sag conditions. Furthermore, owing to the structurally decoupled voltage-control framework, the recovery boundary becomes less sensitive to current-angle variation and transient voltage-controller dynamics. Consequently, recovery consistency is improved and oscillatory CLC-CVC transitions are suppressed. Once the synchronization trajectory re-enters the recoverable operating region and the demanded current falls below the allowable limit, the controller smoothly returns to normal CVC operation.
\subsection{Dynamic Recovery Characteristics of the Proposed Strategy}

\begin{figure}[!t]
	\centering
	\includegraphics[width=2.4 in]{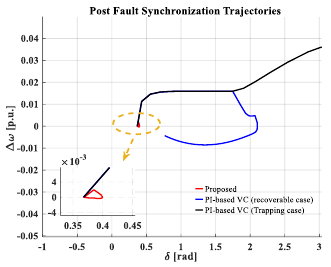}
	\caption{Time-domain simulation results of the reduced-order synchronization model comparing PI-based voltage control and the proposed structurally decoupled recovery framework.}
	\label{fig:phase_portrait}
\end{figure}
To illustrate the analytical observations, time domain simulations of the reduced-order synchronization model are performed under current-limited operation. The resulting synchronization trajectories are compared in Fig.~\ref{fig:phase_portrait}. Under conventional PI-based voltage control, the synchronization trajectories exhibit comparatively larger angle and frequency excursions, and may either converge to the stable equilibrium region or diverge beyond the unstable equilibrium boundary depending on the recovery condition.

In contrast, the proposed framework produces more compact and convergent trajectories with reduced synchronization-angle excursion and frequency deviation. This improvement results from the combined effect of candidate-based current-angle steering and structural decoupling between the voltage controller and the current limiter. Consequently, smoother transition from CLC to CVC operation is achieved, while trajectory trapping and oscillatory recovery behavior are suppressed.

\section{Experimental Validation}

To validate the proposed fault recovery enhancement framework, experiments are conducted on a laboratory scale grid-connected GFM inverter under severe voltage sag conditions. The investigation focuses on synchronization recovery during current-limited operation, with emphasis on trapping behavior, CLC-CVC transitions, and the influence of the voltage control structure on recovery dynamics.

\subsection{Experimental Setup}

The proposed strategy is experimentally validated using a 3-kVA laboratory-scale grid-connected GFM inverter prototype. The inverter is implemented as a two-level voltage source converter supplied by a $300~\mathrm{V}$ dc link, with a nominal phase voltage of $122.5~\mathrm{V}$ (peak). The active power reference is set to $2.4~\mathrm{kW}$ ($0.8~\mathrm{p.u.}$), while the converter current is limited to $1.2~\mathrm{p.u.}$ The system operates under weak-grid conditions corresponding to a short-circuit ratio (SCR) of approximately 4.0. The control algorithm is implemented on an IMPERIX real-time control platform using a B-Box RCP interfaced with a PEB 8038 power electronic board.

\begin{figure}[!t]
	\centering
	\includegraphics[width=3.5in,height=1.4in]{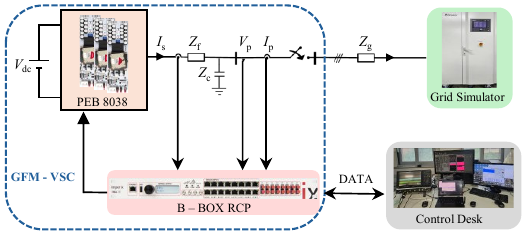}
	\caption{Experimental setup of the laboratory scale GFM inverter platform.}
	\label{fig:exp_setup}
\end{figure}

The dc-link source and programmable grid conditions are emulated using a bidirectional dc power supply (IT6018T) and a programmable grid simulator (ANBGS), respectively. The proposed control structure is implemented experimentally. Fig.~\ref{fig:exp_setup} shows the laboratory platform, Table~\ref{tab:parameters} lists the main system and controller parameters, and Table~\ref{tab:grid_sensitivity} evaluates grid-impedance sensitivity.

\begin{table}[!t]
	\caption{System and Control Parameters}
	\label{tab:parameters}
	\centering
	\footnotesize
	\renewcommand{\arraystretch}{1.1}
	\setlength{\tabcolsep}{7pt}
	
	\begin{tabular}{|l|c|}
		\hline
		Parameter & Value \\
		\hline
		Rated power & 3 kVA \\
		DC-link voltage & 300 V \\
		Grid phase voltage & \(86.6\sqrt{2}\) V \\
		Grid angular frequency & \(120\pi\) rad/s \\
		Filter inductance \(L_f\) & 2 mH \\
		Filter capacitance \(C_f\) & \(10~\mu\mathrm{F}\) \\
		Grid inductance \(L_g\) & 5 mH \\
		Grid resistance \(R_g\) & \(0.2~\Omega\) \\
		Maximum current \(I_{s,\max}\) & 1.2 p.u. \\
		Switching frequency & 20 kHz \\
		Sampling frequency & 40 kHz \\
		Steering weights \((\lambda,\rho)\) & \((8,0.2)\) \\
		Steering thresholds \((s_{\mathrm{low}},s_{\mathrm{crit}},V_{\mathrm{deep}})\) & \((0.02,0.05,0.60)\) \\
		Candidate settings \((N_{\mathrm{cand}},\Delta_{\mathrm{SEP}})\) & \((11,0.15~\mathrm{rad})\) \\
		\hline
	\end{tabular}
\end{table}

\subsection{Conventional Fixed-Angle Current-Limiting Performance}

\begin{figure}[!t]
	\centering
	
	\subfloat[]{
		\includegraphics[width=3.4in,height=1.4in]{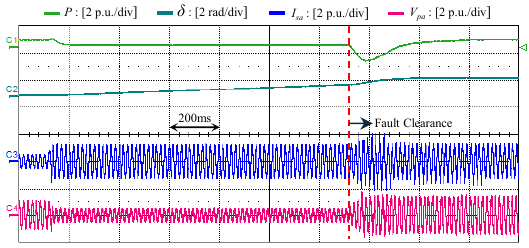}
		\label{fig:trap_case}
	}
	
	\vspace{1mm}
	
	\subfloat[]{
		\includegraphics[width=3.4in,height=1.4in]{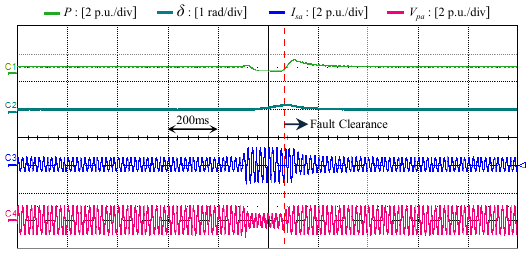}
		\label{fig:recovery_case}
	}
	
	\vspace{1mm}
	
	\subfloat[]{
		\includegraphics[width=3.4in,height=1.4in]{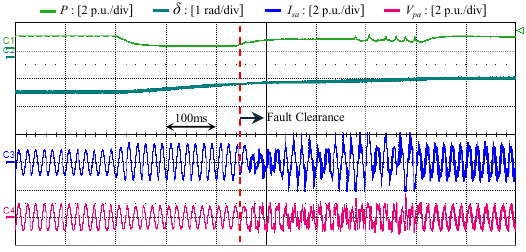}
		\label{fig:oscillation_case}
	}
	
	\caption{Experimental responses under conventional fixed angle current limiting. 
		(a) Converter remains trapped in CLC after fault clearance. 
		(b) Successful recovery from CLC to CVC after fault clearance. 
		(c) Oscillatory transitions occur between CLC and CVC during recovery.}
	
	\label{fig:conv_cases}
\end{figure}

Experiments are first conducted using conventional fixed angle current limiting together with PI-based voltage regulation to investigate the influence of the injected current-angle on post-fault synchronization behavior. In Figs.~\ref{fig:conv_cases} and~\ref{fig:proposed_cases}, 
the CLC period corresponds to the interval during which the current magnitude $I_{sa}$ is saturated at $1.2$~p.u., while CVC operation is identified by the restoration of 
the PCC voltage $V_{pa}$ to its nominal value following fault clearance.

A three phase to ground voltage sag is applied to trigger current-limited operation. Fig.~\ref{fig:conv_cases}(a) shows the experimental response for a fault duration of \(1.2~\mathrm{s}\) with a fixed injected current-angle of \(\phi=-0.2~\mathrm{rad}\). Although synchronization is initially maintained during the disturbance, the synchronization-angle continuously increases after fault clearance and the converter remains trapped in CLC without returning to normal voltage regulation.

When the fault duration is reduced to \(0.15~\mathrm{s}\) while maintaining the same injected current-angle, the converter successfully returns to CVC operation after fault clearance, as shown in Fig.~\ref{fig:conv_cases}(b). Under this condition, the synchronization trajectory remains within the recoverable operating region and stable post-fault recovery is achieved.

To further illustrate the influence of the injected current-angle, the current-angle is changed to \(\phi=-1.4~\mathrm{rad}\) under a fault duration of \(0.25~\mathrm{s}\). As shown in Fig.~\ref{fig:conv_cases}(c), the converter exhibits repeated oscillatory transitions between CLC and CVC during fault recovery, preventing convergence to a stable post-fault operating condition.

These results indicate that post-fault synchronization recovery under conventional fixed angle current limiting is highly sensitive to both fault conditions and injected current-angle. Although appropriate current-angle selection may improve the restoring synchronization characteristic, stable recovery cannot be consistently guaranteed using fixed angle current limiting alone.
\subsection{Influence of Voltage Control Structure and Fault Robustness}

To further investigate the interaction between current limitation and voltage control dynamics, the proposed candidate-based current-angle steering strategy is first implemented together with conventional PI-based voltage regulation.

\begin{figure*}[!t]
	\centering
	
	\subfloat[]{
		\includegraphics[width=3.4in,height=1.4in]{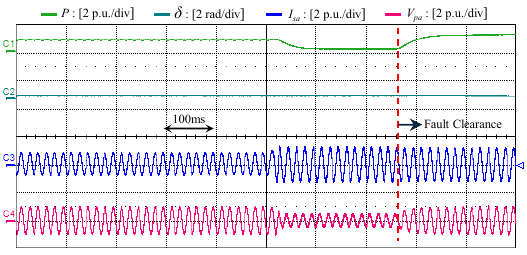}
		\label{fig:pi_candidate}
	}
	\hfil
	\subfloat[]{
		\includegraphics[width=3.4in,height=1.4in]{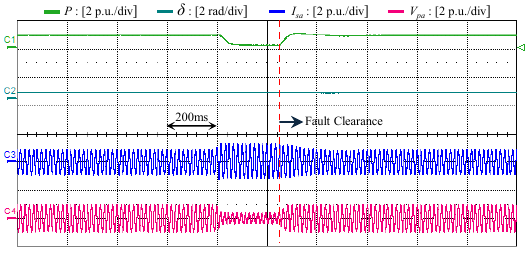}
		\label{fig:proposed_recovery}
	}
	
	\vspace{1mm}
	
	\subfloat[]{
		\includegraphics[width=3.4in,height=1.4in]{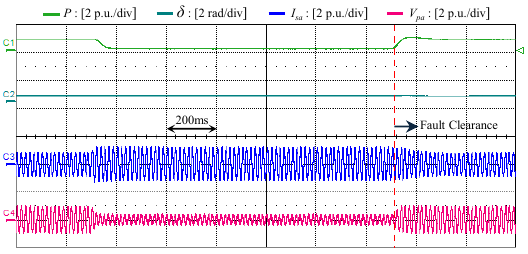}
		\label{fig:proposed_long_fault}
	}
	\hfil
	\subfloat[]{
		\includegraphics[width=3.4in,height=1.4in]{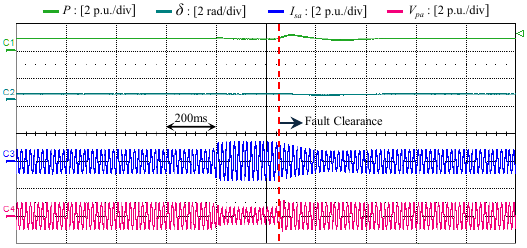}
		\label{fig:slg_recovery}
	}
	
	\caption{Experimental validation of the proposed recovery framework. 
		(a) Candidate-based steering with PI-based voltage 
		regulation incorporating anti-windup compensation. 
		(b) Proposed virtual admittance based voltage regulation. 
		(c) Stable recovery under prolonged fault duration. 
		(d) Recovery performance under SLG fault condition.}
	
	\label{fig:proposed_cases}
\end{figure*}

As shown in Fig.~\ref{fig:proposed_cases}(a), trapping 
behavior persists even when candidate-based current-angle steering is combined with PI-based voltage regulation 
incorporating anti-windup compensation (integrator frozen 
during current saturation). This result demonstrates that 
post-fault recovery failure is not solely attributable to 
integrator windup, but arises from the structural coupling 
between the voltage controller dynamics and the current 
limiter, which persists even after windup is suppressed.

The proposed virtual admittance based voltage control 
structure is subsequently applied together with the 
candidate-based steering strategy. Under the same 
operating condition, the converter successfully returns 
to CVC operation without trapping or oscillatory 
transitions, as shown in Fig.~\ref{fig:proposed_cases}(b).

To further validate robustness, the fault duration is 
increased to $1.2~\mathrm{s}$. As shown in 
Fig.~\ref{fig:proposed_cases}(c), the converter still 
achieves stable post-fault recovery without trapping or 
oscillatory behavior despite the prolonged disturbance.

The robustness of the proposed framework under unsymmetrical disturbances is further examined using an SLG fault under the same weak grid operating condition. As shown in Fig.~\ref{fig:proposed_cases}(d), the converter successfully restores CVC operation after fault clearance without trapping or sustained CLC-CVC oscillation, confirming effective operation under unsymmetrical voltage sag conditions.

\begin{table}[!t]
	\caption{Sensitivity to Grid Impedance Variation With \(R_g=0.2~\Omega\)}
	\label{tab:grid_sensitivity}
	\centering
	\footnotesize
	\renewcommand{\arraystretch}{1.2}
	\setlength{\tabcolsep}{5pt}
	
	\begin{tabular}{|c|c|c|}
		\hline
		\(L_g\) & Max. \(\delta\) Excursion & Max. \(\Delta\omega\) (p.u.) \\
		\hline
		2 mH  & 0.27 rad & \(1.9\times10^{-3}\) p.u. \\
		\hline
		5 mH  & 0.39 rad & \(2.0\times10^{-3}\) p.u. \\
		\hline
		8 mH  & 0.52 rad & \(2.1\times10^{-3}\) p.u. \\
		\hline
		10 mH & 0.61 rad & \(2.1\times10^{-3}\) p.u. \\
		\hline
	\end{tabular}
\end{table}

\subsection{Discussion of Experimental Results}

The experimental results confirm that post-fault recovery of current-limited GFM converters is governed by both synchronization trajectory evolution and recovery boundary dynamics. Under conventional PI-based voltage regulation, trapping and oscillatory CLC-CVC transitions persist even with candidate-based current-angle steering, indicating that favorable current-angle selection alone is insufficient to guarantee stable recovery. In contrast, the proposed virtual admittance based voltage control structure maintains structural decoupling between the unconstrained soft current and the saturated converter current, reducing the sensitivity of the recovery boundary to current-angle variation and transient controller dynamics. Consequently, the proposed framework achieves reliable post-fault recovery by combining structural decoupling of the recovery boundary with current-angle-based trajectory regulation. Additional sensitivity studies summarized in Table~\ref{tab:grid_sensitivity} confirm robust recovery across the tested range of grid inductances without controller retuning.

\section{Conclusion} 
This paper investigated the post-fault recovery dynamics of current-limited grid-forming converters, with emphasis on the interaction between the current limiter and the voltage control structure. The analysis showed that conventional PI-based voltage regulation introduces a moving switching boundary mechanism during current-limited operation, which may lead to trapping in CLC or oscillatory CLC-CVC transitions during post-fault recovery. 

To address these issues, a fault recovery enhancement framework combining candidate-based current-angle steering and structurally decoupled virtual admittance based voltage regulation was proposed. The proposed framework provides a practical approach for improving post-fault synchronization recovery of current-limited GFM converters under weak-grid conditions while preserving implementation simplicity. Experimental validation on a 3-kVA laboratory scale GFM inverter under symmetrical and unsymmetrical voltage sag conditions verified that the proposed framework effectively suppresses trapping and oscillatory recovery behavior while improving post-fault synchronization recovery under severe disturbances. These results demonstrate the effectiveness of structural decoupling and current-angle steering for post-fault recovery of current-limited GFM converters.

\end{document}